\documentclass[12pt,a4paper]{article}

\usepackage[T1]{fontenc}
\usepackage[hmarginratio=1:1,top=32mm,columnsep=20pt]{geometry}
\usepackage{mathdots}

\usepackage{color}
\definecolor{dark-gray}{gray}{0.20}
\definecolor{gray}{gray}{0.30}
\definecolor{light-gray}{gray}{0.80}
\definecolor{dark-red}{rgb}{0.7,0,0}
\definecolor{dark-green}{rgb}{0.1,0.4,0}
\definecolor{dark-blue}{rgb}{0.3,0.3,0.7}
\definecolor{light-blue}{rgb}{0.8,0.8,1}

\usepackage{cite}
\usepackage{hyperref}
\hypersetup{
	colorlinks=true,
	linkcolor=dark-blue,
	citecolor=dark-red,
	urlcolor=dark-green,
	linktoc=page
}

  \usepackage{a4wide}
  \usepackage{latexsym}
  \usepackage{epsf}
  \usepackage{bbm}
  \usepackage{graphicx}
  \usepackage{amsmath,amssymb,amsthm}
  \usepackage{verbatim}

\newcommand{\bbm}{\left(\begin{matrix}}
\newcommand{\ebm}{\end{matrix}\right)}
\newcommand{\bea}{\begin{eqnarray}}
\newcommand{\eea}{\end{eqnarray}}
\newcommand{\be}{\begin{equation}}
\newcommand{\ee}{\end{equation}}

\renewcommand{\cal}[1]{\mathcal{#1}}

\begin{document}

\numberwithin{equation}{section}

\begin{center}

{\LARGE {\bf Rigid rotation in GR and a generalization of the virial theorem for gravitomagnetism }}  \\

\vspace{1.5 cm} {\large  Davide Astesiano$^a$ }\footnote{{ \upshape\ttfamily dastesiano@uninsubria.it } }\\
\vspace{1 cm}  ${}^a$ Department of Science and High Technology, Università dell'Insubria,\\
Via Valleggio 11, 22100, Como, Italy
$\&$ INFN, sezione di Milano\\ Via Celoria 16, 20133, Milano, Italy

\vspace{2cm}

{\bf Abstract}
\end{center}

{\small In this work we study the properties of rigidly rotating neutral dust solutions in general relativity. This class of solutions gained relevance recently due to applications to the dynamics of spiral galaxies. We show that this class could be interpreted as a “rigid body" in general relativity and we analyze the different properties respect to the rigidly rotating disk in special relativity: for example, the general relativistic counterpart shows no Doppler effect for a light signal emitted and received from any two points at rest respect to the “rigid body". This effect can be important to test the validity of the assumed model for our galaxy.\\
In the second part we approach the problem from a low energy expansion perspective and we write down a generalization of the virial theorem for stationary spacetimes. The non-Newtonian contributions can lead to a re-weighting of dark matter in galaxies.}

\setcounter{tocdepth}{2}
\newpage

\textcolor{white}{x}\\

\section{Introduction}
\noindent
In the work \cite{Rosen1947NotesOR} a tensor of deformations $\tilde{P}$ is introduced to define covariant equations for “rigid-body" motions in special relativity. Using this definition it is possible to show that a body which is rigidly rotating around an axis satisfies the ``linear speed-distance law" (\ref{omega1}). We will use a different definition for the deformation tensor respect to \cite{Rosen1947NotesOR}, eventually reaching same conclusions. In the first part we review this construction in order to study a generalization of the definition of  rigid rotation around an axis in a general relativistic context. To do this, we use dust as source for the Einstein's equations
\begin{align}
    T= \rho\, u \otimes u,
\end{align}
where $u$ is the four velocity field of the rotating body and $\rho$ its density. \\
The class of solutions under consideration has gained relevance recently due to applications to spiral galaxy models \cite{Cooperstock:2005qw}, \cite{Balasin:2006cg},\cite{Crosta:2018var}. Our analysis shows that these solutions are rigidly rotating even if they do not satisfy $r^{-1} v=$ constant, where the velocity $v$ is measured by the “zero angular momentum observers" (ZAMO) \cite{1972ApJ}. \\
In the second part we consider again dust solutions in general relativity, without imposing the condition of rigid rotation. As well as writing down exact solutions, one can describe these systems taking a gravito-magnetic expansion approach as in \cite{Ludwig:2021kea} and \cite{Ruggiero:2021lpf}. In fact, in the second part we will approach stationary solutions from a “low energy" expansion perspective and we will provide a generalization of the Newtonian virial theorem in this general relativistic context,which can be useful for galactic and extra-galactic applications. Interestingly the relativistic modification of the theorem presents an extra term which can reduce the required matter density to sustain the motion of the dust compared to the Newtonian version of the same theorem.\\
In what follows we take $c=1, G=1$ and the signature is mostly plus.
\section{Considerations on rigidly rotating dust}
\noindent
Here we consider a rotating disk around an axis in special relativity, therefore neglecting its gravitational field.\\
Given a coordinate system $x^{\mu}$, in \cite{Rosen1947NotesOR} the tensor of deformations $\tilde{P}$ is defined as
\begin{align} \label{Rosen}
    \tilde{P}_{\mu\nu}:= \frac{1}{2}\left(u_{\mu;\nu}+u_{\nu;\mu}+u_{\mu;\alpha} u^{\alpha} u_{\nu}+u_{\nu;\alpha} u^{\alpha} u_{\mu} \right),
\end{align}
where $u^\mu$ is the four velocity field of the body. The condition for rigid rotation adopted in \cite{Rosen1947NotesOR} is $\tilde{P}_{\mu\nu}=0$:
\begin{align}
    \left(u_{\mu;\nu}+u_{\nu;\mu}+u_{\mu;\alpha} u^{\alpha} u_{\nu}+u_{\nu;\alpha} u^{\alpha} u_{\mu} \right)=0. \label{RBC}
\end{align}
We are here taking a different perspective, eventually reaching the same conclusions. We define the tensor of deformations $P$ as 
\begin{align}
    & P(u) := \mathcal{L}_u(g)= (u_{\mu;\nu} + u_{\nu;\mu}) dx^\mu \otimes dx^\nu  \label{deformation},
\end{align}
and its representation in covariant components read
\begin{align}
   P_{\mu\nu}=u_{\mu;\nu} + u_{\nu;\mu}. \label{Deformation1}
\end{align}
If we now directly impose the nullity of $P$
\begin{align}
    u_{\mu;\nu}+u_{\nu;\mu}=0, \label{KillingCond}
\end{align}
we obtain the Killing vector equations. If we want $u$ to be the four-velocity of a real element of the body, then $(\ref{KillingCond})$ must be supplemented by the condition $u^\mu u_\mu=-1$. In fact, the combination of these two set of equations give the geodesic equation
\begin{align}
    u^\mu u^\nu_{;\mu}=0.
\end{align}
The class of solutions for $P=0$, plus the obvious condition $u^\mu u_\mu=-1$, is a family of particles moving on straight lines. Although this is clearly a rigid body motion, the condition $P=0$ is too strong and does not allow for rigid rotation. \\
Let us now take a step back and consider the vector field given by the four velocity of the body
\begin{align}
  u=  u^\mu \frac{\partial}{\partial{x^{\mu}}}.
\end{align}
As it is well known, this vector field naturally defines a projection operator on tensor fields at every point $x$. Its action on a vector $X$ is given by
\begin{align}
    \mathcal{P}_{t}(X):= -u^{\mu}X_{\mu} , \quad  \mathcal{P}_{s}(X):= X+\mathcal{P}_{t}(X).
\end{align}
For obvious reasons we can call $P_{t}(X)$ and $P_{s}(X)$ respectively the temporal and the spatial projections of the vector field $X$ respect to $u$. This action can be trivially extended on tensors with two indices if we clearly state which index we are acting upon. In this sense, since we take under consideration the deformation tensor field $P_{\mu\nu}$ in eq.(\ref{Deformation1}) which is symmetric, the action of $u$ is
\begin{align}
   \mathcal{P}_{t}\mathcal{P}_{t}(P)&:= u^{\mu} P_{\mu\nu} u^{\nu}=0, \\
   \mathcal{P}_{t}(P)&:=u^{\mu} P_{\mu\nu}= u^\mu u^\nu_{;\mu}.
\end{align}
Therefore we can define the “spatial part" of the deformation tensor $P$ the tensor
\begin{align}
    \frac{1}{2}\left(u_{\mu;\nu}+u_{\nu;\mu}+u_{\mu;\alpha} u^{\alpha} u_{\nu}+u_{\nu;\alpha} u^{\alpha} u_{\mu} \right),
\end{align}
which is exactly $\tilde{P}_{\mu\nu}$ in eq.(\ref{Rosen}). The term “spatial part" clearly refers to the fact that it is orthogonal to $u$, in fact
\begin{align}
    u^{\mu} \tilde{P}_{\mu\nu}=0,
\end{align}
holds identically. The condition for rigid motion adopted in \cite{Rosen1947NotesOR} is equivalent to require that the projection perpendicular to $u$ of the deformation tensor $P$ is zero, i.e. $\tilde{P}_{\mu\nu}=0$. From our perspective we are imposing the rigid body condition on the space orthogonal to the four-velocity $u$ of the elements of the rigid body.\\ 
In an inertial frame using cylindrical coordinates $(t,r,z,\phi)$, the circular motion of a particle around the $\phi$-axis can be described as
\begin{align}
   u^0=\sqrt{1+\sigma^2r^2},\qquad u^{x}= -\sigma y, \qquad u^{y}=\sigma x,
\end{align}
where $\sigma=\sigma(r)$ and $r^2=x^2+y^2$. As explained in \cite{Rosen1947NotesOR}, inserting this into eq.(\ref{RBC}) we get 
\begin{align}
    \frac{d\sigma}{dr}=\sigma^3 r,
\end{align}
which has for its solution
\begin{align}
    \sigma= \frac{\omega}{\sqrt{1-\omega^2r^2}},
\end{align}
provided that $\omega$ is a constant. We can get the three dimensional velocity $v$ using
\begin{align}
  v \gamma:= \sqrt{ (u^1)^2+(u^2)^2}= \sigma r, 
\end{align}
which gives the usual Newtonian relation
\begin{align}
    v=\omega r, \label{omega1}
\end{align}
to which before we referred to as the “linear speed-distance law".\\
The coordinates transformation from the inertial reference frame in cylindrical coordinates to the rigidly rotating frame is given by
\begin{align}
\phi \rightarrow \phi +\omega t, \label{RRF}
\end{align}
while the other coordinates remain the same. As it is well known, we arrive at the Born line element
\begin{align}
    ds^2= -(1-\omega^2 r^2) dt^2 +2 \omega r^2 dt d\phi + r^2 d\phi^2+ dr^2 +dz^2,  \label{R}
\end{align}
in what follows will be useful to interpret this as a stationary gravitational field $\Phi_C=-\frac{\omega^2}{2} r^2 $.\\
A useful class of observers in the following discussion are the “zero angular momentum observers" (ZAMO) \cite{1972ApJ}, \cite{Bini:2008uyx}. They are orthogonal to the constant time spacelike hypersurfaces and their orthonormal frame in the case we are considering here, described in terms of the coordinates in (\ref{R}) is
\begin{align}
    e^{(t)}=dt,\quad e^{(\phi)}=d\phi+\omega dt,\quad e^{(r)}= dr,\quad e^{(z)}=dz.  \label{VZ}
\end{align}
Of course, in this specific case the ZAMO is in every point the inertial reference frame, which is a rotating frame as seen from the rigidly rotating frame defined by (\ref{RRF}) with angular velocity $\omega$. In fact, the inverse of the transformation $(\ref{RRF})$ sends back the frame $(\ref{VZ})$ of the ZAMO to
\begin{align}
    e^{(t)}=dt,\quad e^{(\phi)}=d\phi,\quad e^{(r)}= dr,\quad e^{(z)}=dz.
\end{align}
The rigidly rotating metric $(\ref{R})$ written on the basis of the orthonormal frame of the ZAMO $(\ref{VZ})$ is
\begin{align}
    ds^2=- dt^2 + r^2 \left(d\phi+\omega dt\right)^2+ dr^2 + dz^2.  \label{ZAMO1}
\end{align}
\subsection{Rigid rotation in General Relativity}
Now we pass to consider the same problem in a general relativistic context, without neglecting the energy-density of the matter. We will be a little more general and consider neutral stationary, axysimmetric dust coupled to Einstein equations. The matter is assumed to flow along the Killing vectors $\partial_t$ and $\partial_\phi$ and depends only on the coordinates which are not associated to Killing vectors $(r,z)$. Denoting the matter density with $\rho$, in cylindrical coordinates the energy momentum tensor is given by
\begin{align}
    T^{\mu\nu} (r,z)= \rho(r,z) u^{\mu}(r,z) u^{\nu}(r,z), \quad u^{\mu}(r,z)= u^t(r,z) \left(1,0,0,\Omega(r,z) \right),
\end{align}
where $\Omega(r,z)= \frac{d\phi}{dt}= \frac{u^{\phi}}{u^{t}}$. Here we are interested in the case of constant angular velocity $\Omega(r,z)=\Omega$. For constant $\Omega$, we can perform a rigid rotation of the coordinates to rewrite the four velocity of the dust as
\begin{align}
    u= \partial_t.
\end{align}
The dust is now at rest compared to these new coordinate set and the class of solution is now given by \cite{Winicour}
\begin{gather}
    ds^2 = - \left(dt-\eta d\phi\right)^2+ r^2 d\phi^2+e^{\mu}\left(dr^2+dz^2\right),  \label{RRD}\\
    \eta_{,rr}+\eta_{,zz}-\frac{\eta_{,r}}{r}=0,\quad \mu_{,r}= \frac{(\eta_{,z})^2-(\eta_{,r})^2}{2r}, \quad \mu_{,z}=- \frac{\eta_{,r}\eta_{,z}}{r},  \label{RRD1}
\end{gather}
where we used the notation $\partial / (\partial a) (f):= f_{,a}$ with $a=r,z$.\\
To compare this metric with the metric of the rigidly rotating frame (\ref{ZAMO1}) we must rewrite $(\ref{RRD})$ in terms of the vierbein of the ZAMO. The ZAMO's orthonormal frame in this case is
\begin{align}
    e^{(t)}=\frac{1}{\sqrt{1-v^2}} dt, \quad  e^{(\phi)}=r \sqrt{(1-v^2)} \left(d\phi-\chi dt\right),\quad e^{(r)}= e^{\mu/2} dr, \quad e^{(z)}= e^{\mu/2} dz,\end{align}
where
\begin{align}
    \chi= -\frac{v}{r} \left(\frac{1}{1-v^2}\right),\quad v(r,z);= \frac{\eta(r,z)}{r} \label{CC}
\end{align}
The function $v(r,z)$ is the velocity of the dust as judged from the ZAMO, defined by
\begin{align}
    -e^0_\mu u^\mu:=\frac{1}{\sqrt{1-v^2}}.
\end{align}
Note that the definition of ZAMO consistently satisfies the requirement of zero angular momentum $ g_{\phi\phi}\, \chi+g_{\phi t}=0$. The rotating dust metric (\ref{RRD}) written in the basis of the vierbein of the ZAMO is
\begin{align}
   ds^2=- \frac{1}{1-v^2} dt^2 + r^2(1-v^2) \left(d\phi-\chi dt\right)^2+ dr^2+dz^2.  \label{ZAMO2}
\end{align}
We now want to show that in the limit of $\rho \rightarrow 0$, this class of solutions approach the rigidly rotating disk solution. The energy density $\rho(r,z)$ is given by \cite{Winicour}
\begin{align}
    8 \pi G \rho e^{\mu}=  \frac{(\eta_{,r})^2+(\eta_{,z})^2}{r^2},  \label{density}
\end{align}
therefore we are interested in taking the limit for little $v$. If we expand the metric (\ref{ZAMO2}) in powers of $v$ we see that the expansion up to order $v$ is exactly the metric of the rigidly rotating frame (\ref{ZAMO1}).
This reflects the fact that we are describing the general relativistic solution in the coordinates of a rigidly rotating frame. This rigidly rotating frame is at rest compared to the dust and it is in free falling with it.\\
From the metric (\ref{ZAMO2}) written in terms of the vierbein of the ZAMO,  we can read the gravitational potential, which is
\begin{align}
    \Phi_N= \frac{v^2}{2},
\end{align}
and gives the heuristic interpretation of the solution: the gravitational potential due to the presence of the dust exactly balance the gravitational potential of the non inertial force given by the rigidly rotation of the reference frame. In fact the centrifugal force in the rigidly rotating disk, with angular velocity $\chi$, is $\Phi_C= -\frac{\chi^2}{2}r^2$ and then we have the usual balance equation
\begin{align}
    \Phi_N+ \Phi_C=\frac{v^2}{2}-\frac{\chi^2}{2}r^2 = 0+O(v^{2}), \label{EE}
\end{align}
where we used eq.(\ref{CC}) up to order $v$
\begin{align}
    \chi= -\frac{v}{r}+ O(v^{2}).
\end{align}
The compensation of the two stationary gravitational forces is also the reason why $g_{tt}=-1$ everywhere in the full solution (\ref{RRD}), which allows us to choose a preferred time. \\
This class of rotating dust solutions has interesting properties, now we investigate the structure in more detail.\\
The dust has four velocity
\begin{align}
    u= \partial_t,
\end{align}
obviously, being $\partial_t$ a Killing vector eq.(\ref{KillingCond}) is satisfied. Therefore, all the components of the tensor of deformations vanish:
\begin{align}
    P=0.
\end{align}
This condition shows again that the dust is in rigid rotation even though $r^{-1} v$ is not constant in general.
The condition $P=0$ is stronger than the analog condition for the rigidly rotating disk in special relativity, for which only the “spatial part" $\tilde{P}$ of the tensor of deformations is zero. \\
The rigid rotation (actually the nullity of $P$) has another interesting implication for astrophysical measurements, since it allows for the existence of a timelike Killing vector for which $g_{tt}=g_{\mu\nu} (\partial_t)^\mu(\partial_t)^\nu=-1$. Let us imagine that a light signal is emitted from a generic element of the body and it is received from another element of the body, the measured redshift (or blueshift) is 0, there is no shift in the frequence of the light! The reason is that for every element composing the body the measured frequency $\nu$ is
\begin{align}
    k^\mu u_{\mu}= -\nu
\end{align}
but $u= \partial_t$ is the same for every observer at rest respect to the rigid body. Again this reflects the compensation in eq.(\ref{EE}).
The same experiment performed on a rigidly rotating disk in special relativity, where the metric is defined by eq. (\ref{R}), would yield the following result for the frequency shift between two the emitter (E) and the receiver (R) \cite{Synge}
\begin{align}
    \frac{\nu_E}{\nu_R}= \frac{\sqrt{1- \omega^2 r^2_R}}{\sqrt{1- \omega^2 r^2_E}},
\end{align}
then it is different from 1 if they are at different radious.\\
The nullity of the light frequency shift can be important for applications to real galaxies. If we are co-rotating with a rigidly rotating galaxy and we try to apply the special relativistic formula of the doppler shift, therefore neglecting the effect of gravity, the results would be completely wrong. There is no shift between a light signal emitted and received at any two points of the rigid body and this effect does not depend on the regime of application. \\
The works \cite{Balasin:2006cg} and \cite{Crosta:2018var} propose this class of rigidly rotating solutions as a model for our galaxy. The validity of the model can be confirmed if the measurements actually reproduce the nullity of the light frequency shift from the other stars in our galaxy. 

\section{Low energy expansion and a virial theorem for gravitomagnetism} \label{NPA}
\noindent
In this section we reintroduce the constants $c$ and $G$. We want to study the system of stationary neutral dust coupled to Einstein equations as in the last section, but now we address the problem starting from a low energy expansion. Then, as the standard procedure, we assume the existence of global coordinates $(t,x^i)$ in which the metric reads
\begin{gather}
    g_{\mu\nu}=\eta_{\mu\nu}+h_{\mu\nu}, \qquad |h_{\mu\nu}| \ll 1. \label{g1}
\end{gather}
These coordinates select a preferred reference frame $\mathcal{I}$, defined by the family of curves $t=$variable. In this reference frame all the matter content is slow compared to the speed of light. To do this, we perform an expansion in power of $1/c$ and we will consider the leading orders. We write the solution of Einstein field equations in weak-field and slow-motion approximation exploiting a well known analogy with Maxwell equations: this is the so-called \textit{gravitoelectromagnetic} formalism (see e.g. \cite{Ruggiero:2002hz,Mashhoon:2003ax}); accordingly, the line element describing this solution is
\begin{align}
    ds^2= -c^2 \left(1-2\frac{\Phi}{c^2}\right) dt^2 -\frac4c A_{\hat{i}} dx^{\hat{i}} dt + \left(1+2\frac{\Phi}{c^2}\right)\delta_{ij}dx^{\hat{i}} dx^{\hat{j}}. \label{eq:weakfieldmetric1}
\end{align}
The coordinates are splitted as $\{\hat{i},\hat{j}\}= \{x,y,z\}$ are the spatial indices. Accordingly, we must choose a form of the energy momentum tensor suitable for such expansion of the geometry. For the four-velocities and the energy-momentum tensor of matter $T^{\mu\nu}$ we assume the following approximation
\begin{align}
    u^0=&1+ O(c^{-2}), \quad u^{\hat{i}}= v_{\hat{j}} + O(c^{-3}),\\
    T^{00}=&\rho+ O(\rho c^{-2}), \quad T^{0\hat{i}}=\rho v_{\hat{j}}+ O(\rho c^{-3}),\quad T^{\hat{j}\hat{k}}= \rho v_{\hat{j}} v_{\hat{k}}+ O(\rho c^{-4}).\label{Tjk}
\end{align} 
In the above equation the gravitoelectric  ($\Phi $) and gravitomagnetic  ($A_i$) potentials, in stationary conditions, are solutions of Poisson equations 
\begin{eqnarray}
\nabla^{2} \Phi&=&-4\pi G\, \rho, \label{eq:poisson01} \\
\nabla\wedge \left( \nabla \wedge  \vec{A} \right) & =& 8\pi G\, \rho \frac{\vec{v}}{c},  \label{eq:poisson02}
\end{eqnarray}
where $\vec{A}$ and $\vec{v}$ clearly refer to the components of the respective vectors. 
Usually in the Newtonian limit, one assume that the dominant contribution to gravitation comes from $T_{00}$ and then neglect the $(0,\hat{i})$ components of the Einstein equations under the assumption $A \sim O(c^{-1}) $.
Introducing
\begin{align}
     B_{\hat{i}}:=\epsilon_{\hat{i}}^{\,\,\hat{j}\hat{k}} \partial_{\hat{j}} A_{\hat{k}}, \label{defb}
\end{align}
allows us to express the time and spatials components of the local conservation equation $T^{\mu\nu}_{\quad;\nu}=0$ respectively as
\begin{gather}
    \frac{\partial \rho}{\partial t}+ \frac{\partial \left(\rho v_{\hat{i}}\right)}{\partial x^{\hat{j}}}=0 ,
\end{gather}
\begin{gather}
    \rho \frac{dv_{\hat{i}}}{dt}=- \rho \left(\Phi_{,\hat{i}}+ 2 \epsilon^{\hat{i}\hat{j}\hat{k}} \frac{v_{\hat{j}}}{c} B_{\hat{k}}\right), \label{meom}
\end{gather}
where we defined $d/dt:= \partial_t+ v_{\hat{j}} \partial_{\hat{j}}$. These equations are well known, the first one is the continuity equation and the second one is the analogue of the Lorentz force law \cite{Ruggiero:2021lpf}.\\
We note that
\begin{align}
    -2 \rho \epsilon^{\hat{i}\hat{j}\hat{k}} \frac{v_{\hat{j}}}{c} B_{\hat{k}}=&- \frac{1}{4\pi G} \epsilon^{\hat{i}\hat{j}\hat{k}} B_{\hat{k}} \epsilon^{\hat{j}\hat{l}\hat{s}} \partial_{\hat{l}} B_{\hat{s}}= \\
    &=\frac{1}{4\pi G} \left( B^{\hat{j}} \partial^{\hat{i}} B_{\hat{j}} -B^{\hat{j}} \partial_{\hat{j}} B^{\hat{i}} \right)= -\frac{1}{4\pi G} \partial_{\hat{j}} \left(B^{\hat{j}}B^{\hat{i}}- \frac{1}{2} \delta^{\hat{i}\hat{j}}B^{\hat{k}} B_{\hat{k}}\right),
\end{align}
where we used the following consequences of eq.(\ref{eq:poisson02}) and eq.(\ref{defb})
\begin{gather}
      \partial_{\hat{j}} \left(\epsilon^{\hat{i}\hat{j}\hat{k}}B_{\hat{k}}\right)=\frac{8\pi G}{c} \rho v^{\hat{i}}, \quad
   B^{\hat{i}}_{\,\,,\hat{i}}=0.
\end{gather}
Thanks to these results it is useful to define the two stress pseudo tensors 
\begin{gather}
    t_{N|\hat{i}\hat{j}}:= \frac{1}{4 \pi G} \left(\Phi_{,\hat{i}}\Phi_{,\hat{j}}- \frac{1}{2} \delta_{\hat{i}\hat{j}} \Phi_{,\hat{k}}\Phi_{,\hat{k}}\right), \\
     t_{\mathcal{J}|\hat{i}\hat{j}}:=  \frac{1}{4 \pi G} \left(B^{\hat{i}} B^{\hat{j}} - \frac{1}{2} \delta^{\hat{i}\hat{j}}B^{\hat{k}} B_{\hat{k}}\right), \label{tj}
\end{gather}
such that $\Phi_{,\hat{i}}= \partial_{\hat{i}} \Phi$. We can rewrite the equation of motion for the matter (\ref{meom}) as
\begin{gather}
    \rho \frac{dv^{\hat{i}}}{dt}= - \partial_{\hat{j}} \left(t_{N|\hat{i}\hat{j}}+t_{\mathcal{J}|\hat{i}\hat{j}}\right),
\end{gather}
or, equivalently
\begin{gather}
  \frac{\partial}{\partial t}  \left(\rho v_{\hat{i}}\right) + \left(t_{N|\hat{i}\hat{j}}+t_{\mathcal{J}|\hat{i}\hat{j}}+ \rho v_{\hat{i}} v_{\hat{j}}\right)_{,\hat{j}}=0.
\end{gather}
At our order of approximation the last equation can be integrated as
\begin{align}
    \frac{d^2 \mathcal{I}_{\hat{i}\hat{j}}}{dt^2}= 2 \int \left( t_{N|\hat{i}\hat{j}}+t_{\mathcal{J}|\hat{i}\hat{j}}+\rho v_i v_j\right)d^3x, \label{VT}
\end{align}
where we used the second moment of the system's mass distribution
\begin{gather}
    \mathcal{I}_{\hat{i}\hat{k}}= \int \rho x_{\hat{i}} x_{\hat{j}} d^3x.
\end{gather}
In this derivation, we made use of the following known result \cite{Misner:1974qy}, \cite{1965ApJ...142.1488C}
\begin{gather}
    \frac{d}{dt} \int \rho f d^3x = \int \rho \frac{df}{dt} d^3x+O(\rho c^{-2}),
\end{gather}
where $f$ is a generic function.Usually one is interested in taking the time ``long" average of $(\ref{VT})$, in the approximation where this average for the second derivative of the second moment of the system's mass distribution is zero
\begin{gather}
    \langle\int t_{N|ij} d^3x+ \int t_{\mathcal{J}|ij} d^3x+ \int \rho v_i v_j d^3x\rangle=0,
\end{gather}
when we take the trace, this equation becomes
\begin{align}
    \langle \int \rho v^2\, d^3x - \frac{1}{2} \int \rho  \Phi\, d^3x +\int t_{\mathcal{J}}\, d^3x\rangle=0,
\end{align}
which can be rewritten in the usual form
\begin{gather}
    \langle 2 \mathcal{K}+ \mathcal{U}- \frac{1}{8 \pi G} \int \mathcal{H} d^3x \rangle=0, \label{VEJ} \\
    \mathcal{H}:= \left(\partial_{\hat{i}} A_{\hat{j}}\right)^2- \left(\partial_{\hat{i}} A_{\hat{j}}\right)\left(\partial_{\hat{j}} A_{\hat{i}}\right). \label{H}
\end{gather}
It is interesting to note that this result can reduce the amount of matter needed to sustain a motion with velocity $v$ compared to the Newtonian version of the same theorem. A formal discussion of the virial theorem for spherical and stationary
axisymmetric spacetimes is given in \cite{Bonazzola1973TheVT}. The approach used here is substantially different and it is not restricted to axisymmetric spacetimes, although we will see an explicit application of (\ref{VEJ}) for axisymmetric systems in the next section. More importantly, the specific chosen expansion allowed for a new explicit result (\ref{VEJ}) that can be directly tested on real galaxies. Moreover, our result makes clear that the off-diagonal terms $h_{0i}$, which have a nice interpretation as  “gravito-magnetic" fields, can have the effect of reducing the energy density content $\rho$ compared to the purely Newtonian case.\\
This modification respect to the Newtonian version of the virial theorem can have interesting applications on systems as galaxies or clusters of galaxies. In particular could lead to a re-evaluation of the amount of required dark matter for such systems. This aspect deserves further investigations. \\
The reduction of the needed amount of matter that arise when considering non negligible off-diagonal terms has been already noted, from an exact solution perspective, in \cite{Astesiano:2021ren}. \\
The virial theorem can be written in a more suggestive way
\begin{align}
    \langle 2\int \rho v^2 d^3x-\frac{1}{8\pi G} \int \left(E^2+B^2\right) d^3x \rangle=0.
\end{align}
Using the analogy with electromagnetism we see that the second term is the total energy stored in the gravity fields. Therefore we have the balance equation
\begin{align}
    2 \times \text{energy of free dust (kinetic energy)} = \text{energy of gravity}.
\end{align}

\subsection{Explicit example in stationary axisymmetric spacetimes} \label{AC}
As an explicit application of the above results, let us take an axisymmetric stationary system dominated by “angular currents"
$h_{0\hat{i}} \equiv S_{\hat{i}}$ which are not time dependent $\partial_t S_{\hat{i}}\sim0$ in our approximation. We then take a simplified version of the metric
\begin{gather}
    ds^2= - c^2 \left(1-2\frac{\Phi}{c^2}\right)  dt^2 -4 \frac{\tilde{S}(r,z)}{c} dt d\phi + \left(1+2\frac{\Phi}{c^2}\right) \left(r^2 d\phi^2 + dr^2+ dz^2\right). \label{MR1}
\end{gather}
We switch to cartesian coordinates $(x,y)$ in the $z=0$ plane
\begin{gather}
    r dr = x dx+y dy, \\
    r^2 d\phi= -y dx+ x dy,
\end{gather}
and one gets the following form for the metric
\begin{align}
    ds^2=&-c^2  \left(1-2\frac{\Phi}{c^2}\right)  dt^2 + 4 \frac{\tilde{S}(r,z)}{c} \frac{y}{x^2+y^2} dt dx -4 \frac{\tilde{S}(r,z)}{c} \frac{x}{x^2+y^2} dt dy+ \nonumber \\
    &+\left(1+2\frac{\Phi}{c^2}\right) \left( dx^2 + dy^2+dz^2 \right). \label{APPM1}
\end{align}
With the conventions of the last section we have the identification 
\begin{align}
    A_{x}= - \tilde{S}(r,z) \frac{y}{x^2+y^2}, \quad A_y=\tilde{S}(r,z) \frac{x}{x^2+y^2}, \quad A_z=0.
\end{align}
 Then we have from eq. (\ref{H})
\begin{gather}
    \mathcal{H}=\frac{\tilde{S}^2_{,z}+\tilde{S}^2_{,r}}{r^2}. \label{DE}
\end{gather}
Assuming reflection symmetry $z \leftrightarrow -z$ at the equatorial axis $z=0$, we can neglect the $z-$derivatives and the result in the equatorial $z=0$ plane is
\begin{align}
    \mathcal{H}_{z=0}= \frac{1}{r^4} \left(x \tilde{S}_{,x} + y \tilde{S}_{,y}\right)^2=  \left(\frac{1}{r^2}\, \vec{r} \cdot \nabla \tilde{S} \right)^2.
\end{align}
We can rewrite the virial equation (\ref{VEJ}) as
\begin{gather}
    \langle \int \rho v^2- \frac{1}{16 \pi G} \int \frac{\tilde{S}^2_{,z}+\tilde{S}^2_{,r}}{r^2}  - \frac{1}{2} \int \rho  \Phi \rangle=0.
\end{gather}
As already noted from the general result (\ref{H}), the off-diagonal term affects the required density. In this particular case of axisymmetric spacetimes the term $\mathcal{H}$ in eq.(\ref{DE}) is always positive and then its presence guarantees the reduction of the needed matter compared to the purely Newtonian case. The resulting difference from the Newtonian version of the theorem is essentially the presence of the term
\begin{gather}
 \frac{1}{r^2} \vec{r} \cdot \nabla \tilde{S}. \label{DJ}
\end{gather}
The same effect can be understood from a different perspective. Let us study the deflection of light due to the presence of an object whose metric can be written as in $(\ref{APPM1})$. We limit the problem to the galactic plane $z=0$, when the angular momentum of the approaching light is different from zero (the impact parameter is $\neq 0$), using an approach similar to \cite{Misner:1974qy}. Here we are neglecting the Newtonian potential $\Phi$ because we want to focus on the effect due to the non-diagonal elements of the metric. We consider the motion in the galactic plane of a photon coming from far away with initial velocity
\begin{gather}
    v^t=\tilde{\omega}, \qquad v^x=\tilde{\omega}, \qquad v^y=0.
\end{gather}
During its trajectory we suppose that $ v^t=\tilde{\omega}$ and $v^x=\tilde{\omega}$ remain approximately constant, while $v^y$ depends only on the value of $x$. The photon arrives from $x=- \infty$, travel across the gravitational field of the galaxy at a nearly constant value of $y=R$, and then flies away at $x=\infty$. During this trip the value of $v^y$ is increased due to the bending of the trajectory by the galaxy, and the deflection angle is
\begin{gather}
    \Delta \phi = \frac{v^y_{Final}}{v^x_{Final}}=\frac{v^y_{Final}}{\tilde{\omega}},
\end{gather}
our approximations are justified by the assumption of a small resulting value for the angle. The geodesic equation is
\begin{gather}
    \frac{d v^y}{d \tau}= \frac{dx}{d\tau} \frac{d v^y}{d x}= \tilde{\omega} \frac{d v^y}{d x}= - (\Gamma^{y}_{tt}+ 2\Gamma^{y}_{tx}+ \Gamma^{y}_{xx}) \tilde{\omega}^2,
\end{gather}
at the first order in $\tilde{S}$, we have
\begin{gather}
    \frac{v^{y}_{Final}|_{y=R}}{\tilde{\omega}}=-  \int^{+\infty}_{-\infty} \frac{\Vec{r} \cdot \vec \nabla \tilde{S} }{r^2} dx |_{y=R}=  -\int^{+\infty}_{-\infty} \frac{y \tilde{S}_{,y}+ x \tilde{S}_{,x}}{x^2+y^2} dx |_{y=R}.
\end{gather}

\section{Final remarks and perspectives}
\noindent
In the first part we studied rigidly rotating neutral dust solutions in general relativity under the assumptions of stationarity and axysimmetry and we tried to understand their properties and their difference with well known solutions in special relativity. 
We discussed the interpretation of the usually adopted coordinates and the nullity of the tensor of deformations $P$ for such solutions. The features of the solutions allowed us to draw some limits of the $\Omega =$ constant model assumed in \cite{Balasin:2006cg} and \cite{Crosta:2018var} as a model for our galaxy. In this sense, we suggested that a strong indicator for the validity of this rigidly rotating model for our galaxy is the nullity of the Doppler effect between any two points (the receiver and the emitter of the light signal) co-rotating with the dust.
This can be important for galactic applications. If this effect is not verified by the observations one should therefore take under consideration the larger class of $\Omega \neq$ constant as a model for our galaxy. The most general solution for rotating dust under the assumptions of stationarity and axysimmetry and its application to galaxies was studied in \cite{Astesiano:2021ren} and more work is in order. \\
In the second part we did not restrict the discussion to rigid rotation and we approached the problem from a different perspective. We generalized the virial theorem to gravito-magnetism and this provided a new formulation that can be tested on real objects. Interestingly, this generalized version of the virial theorem reduces the matter needed to sustain the motion compared to the usually adopted Newtonian version of the same theorem. In the light of the dark matter problem this additional contribution can lead to a re-weighting of the contribution of dark matter in galaxies or in clusters.\\
In future would be interesting to explicitely write down  more general versions of the virial theorem lowering the number of symmetries of the system.

\section*{Acknowledgements}
I would like to thank Professor Matteo Luca Ruggiero for reviewing this manuscript.
\section*{Data availability}
No new data were generated or analysed in support of this research.
\appendix
\section{More details on Gravitomagnetism}
We consider the field of a galactic or an extra-galactic source in the linear approximation of GR. We therefore assumed the metric in the form (\ref{g1}). The Einstein equations can be casted in the form
\begin{align}
\Box \bar{h}_{\mu \nu} &= 16 \frac{\pi G}{c^{4}} T_{\mu \nu},\label{feq}\\
\bar{h}_{\mu
\nu}&=h_{\mu \nu} - {1 \over 2} \eta_{\mu \nu} h,
\end{align}
where $h= g^{\mu\nu}h_{\mu \nu}$ after retaining only the terms linear in $h_{\mu\nu}$. As usual, to obtain the equations in this form, one has to impose the Lorentz gauge condition $\bar{h}^{\mu \nu},_{\nu} = 0$. To obtain a parallelism with the electromagnetic theory, we choose the metric such that $\bar{h}_{00}=4 \Phi / c^{2}$,
$\bar{h}_{0i}=-2A_{i}/c^{2}$ and $\bar{h}_{ij}=O(c^{-4})$, where $\Phi(t,{\bf x})$
is the gravitoelectric potential, ${\bf A}(t,{\bf x})$ is the gravitomagnetic
vector potential and we neglect all terms of order $c^{-4}$ and lower \cite{Ruggiero:2002hz}. From the side of the source,
$T^{00}=c^2 \rho$ is the gravitational “charge" density and
$T^{0i}=c j^{i}$ is the corresponding current.  Thus, far from the source

\begin{equation}
\Phi \sim \frac{GM}{r}, \;\;\;\;\;\;\;\;\;\;\; {\bf A} \sim \frac{G}{c} \frac{{\bf
J} \times {\bf r}}{r^{3}},
\end{equation}

\noindent where $M$ and ${\bf J}$ are the total mass and angular momentum of the
source, respectively.\\
Under these assumptions, the Lorentz gauge condition can be expressed as
\begin{equation}
\frac{1}{c} \frac{\partial \Phi}{\partial t} + \mbox{\boldmath $\nabla$} \cdot
\left( {1\over 2} {\bf A} \right) = 0,
\label{gauge}
\end{equation}
 The spacetime metric involving the gravitoelectromagnetic potentials is then 
given by (\ref{eq:weakfieldmetric1})
\begin{align}
    ds^2= -c^2 \left(1-2\frac{\Phi}{c^2}\right) dt^2 -\frac4c A_{\hat{i}} dx^{\hat{i}} dt + \left(1+2\frac{\Phi}{c^2}\right)\delta_{ij}dx^{\hat{i}} dx^{\hat{j}}. \label{eq:weakfieldmetric2}
\end{align}
At this point we can define the gravito-electro and gravito-magnetic fields
\begin{equation}
{\bf E} =- \mbox{\boldmath $\nabla$} \Phi - {1 \over c} {\partial \over \partial
t} \left( {1\over 2} {\bf A} \right), \;\;\;\;\;\;\;\; {\bf B} = \mbox{\boldmath
$\nabla$}
\times {\bf A}, \label{gem}
\end{equation}

\noindent in close analogy with electrodynamics.  It follows from the field
equations (\ref{feq}) and the gauge condition (\ref{gauge}) that

\begin{equation}
\mbox{\boldmath $\nabla$} \cdot {\bf E} = 4 \pi G \rho,
\label{e1}
\end{equation}

\begin{equation}
\mbox{\boldmath $\nabla$} \cdot \left( {1\over 2} {\bf B} \right) = 0,
\label{e2}
\end{equation}

\begin{equation}
\mbox{\boldmath $\nabla$} \times {\bf E} = - {1\over c} {\partial \over \partial t}
\left( {1 \over 2} {\bf B} \right),
\label{e3}
\end{equation}

\begin{equation}
\mbox{\boldmath $\nabla$} \times \left( {1\over 2} {\bf B} \right) = {1\over
c}{\partial \over \partial t} {\bf E} + {4 \pi \over c} G \:{\bf j}\:,
\label{e4}
\end{equation}
explicitly analogous to the Maxwell equations for the GEM field. \\
The magnetic parts of equations (\ref{e1}) - (\ref{e4})
appear with a factor of $1/2$ as compared to standard electrodynamics
is due to the fact that the effective gravitomagnetic charge
is twice the gravitoelectric charge. \\
Under the assumption of stationarity, from the presented equations one obtains
(\ref{eq:poisson01}) and (\ref{eq:poisson02}).\\
The geodesics of the particles of dust given by (\ref{meom}), can be obtained from the variational principle $\delta \int {\cal L} dt=0$,
where ${\cal L}=mcds/dt$. The lagrangian is
\begin{align}
{\cal L} =&- mc \sqrt{-g_{\mu\nu}\frac{dx^\mu}{dt}\frac{dx^\nu}{dt}}=-mc^2 \sqrt{-g_{00}-2g_{0\hat{i}}\frac{dx^{\hat{i}}}{dt}-g_{\hat{j}\hat{k}} \frac{v^{\hat{j}}v^{\hat{k}}}{c^2}}= \nonumber \\&
- mc^{2} \left[ 1-{v^{2}\over c^{2}} - {2 \over c^{2}} \left( 1+
{v^{2}\over c^{2}} \right) \Phi + {4\over c^{3}} {\bf v} \cdot {\bf A}
\right]^{1/2},
\label{e12}
\end{align}
 which to linear order in $\Phi$ and ${\bf A}$ boils down to

\begin{equation}
{\cal L} = - mc^{2} \left( 1- {v^{2} \over c^{2}} \right)^{1/2} + m \gamma \left(
1+ {v^{2} \over c^{2}} \right) \Phi - {2 m \over c} \gamma {\bf v} \cdot
{\bf A}.
\label{e13}
\end{equation}

The gauge transformations

\begin{equation}
\Phi \rightarrow \Phi - {1\over c} {\partial \psi \over \partial t}, \;\;\;\;\;
{\bf A} \rightarrow {\bf A} +2 \mbox{\boldmath $\nabla$} \psi,
\label{e14}
\end{equation}

\noindent leave the GEM fields (\ref{gem}) and 
hence the GEM equations (\ref{e1})-(\ref{e4})
invariant.  The Lorentz gauge condition (\ref{gauge}) is also satisfied
provided $\Box \psi = 0$. \\
The gravitational field corresponding to the metric (\ref{eq:weakfieldmetric2}) is
given by the Riemann curvature tensor

\begin{equation}
R_{\mu \nu \rho \sigma} = {1\over 2} (h_{\mu \sigma, \; \nu \rho} +
h_{\nu \rho, \; \mu \sigma} - h_{\nu \sigma,\; \mu \rho} - h_{\mu \rho, \nu
\sigma} ),
\label{e15}
\end{equation}

\noindent where
\begin{align}
    h_{00}=2\Phi /c^{2},\quad h_{ij}=(2\Phi /c^{2}) \delta_{ij},\quad h_{0i}=-2A_{i}/c^{2},
\end{align}
plus higher order terms. The components of the Riemann tensor may be expressed in the following form \cite{Mashhoon:1999nr}

\begin{equation}
{\cal R} = \begin{pmatrix} {\cal E} & {\cal B} \\ {\cal B}^{T} & 
{\cal S} \end{pmatrix}\,,
\label{e16}
\end{equation}

\noindent where ${\cal E}$ and ${\cal S}$ are symmetric $3 \times 3$ matrices and
${\cal B}$ is traceless. These matrices read explicitly 

\begin{equation}
{\cal E}_{ij}={1\over c^{2}} E_{j,i}+O(c^{-4}),
\label{e17}
\end{equation}

\begin{equation}
{\cal B}_{ij}=-{1\over c^{2}} B_{j,i} + {1\over c^{3}} \epsilon_{ijk} {\partial
E_{k} \over \partial t} + O(c^{-4}),
\label{e18}
\end{equation}

\noindent and the spatial components are given by

\begin{equation}
{\cal S}_{ij}=-{1\over c^{2}} E_{j,i} + {1 \over c^{2}} (\mbox{\boldmath $\nabla$}
\cdot {\bf E}) \delta_{ij} + O(c^{-4}).
\label{e19}
\end{equation}
They are explicitly invariant under (\ref{e14}). \\
To obtain the generalized virial theorem we defined the tensor $t_{\mathcal{J}|\hat{i}\hat{j}}$ in eq.(\ref{tj}), which again is explicitly invariant under the gauge transformations $(\ref{e14})$, therefore also $\mathcal{H} \sim B^2$ possess the same property.

\bibliographystyle{ieeetr} 
\bibliography{refs} 

\begin{thebibliography}{10}

\bibitem{Rosen1947NotesOR}
N.~Rosen, ``Notes on rotation and rigid bodies in relativity theory,'' {\em
  Physical Review}, vol.~71, pp.~54--58, 1947.

\bibitem{Cooperstock:2005qw}
F.~I. Cooperstock and S.~Tieu, ``{General relativity resolves galactic rotation
  without exotic dark matter},'' 7 2005.
\newblock arXiv:0507619.

\bibitem{Balasin:2006cg}
H.~Balasin and D.~Grumiller, ``{Non-Newtonian behavior in weak field general
  relativity for extended rotating sources},'' {\em Int. J. Mod. Phys. D},
  vol.~17, pp.~475--488, 2008.

\bibitem{Crosta:2018var}
M.~Crosta, M.~Giammaria, M.~G. Lattanzi, and E.~Poggio, ``{On testing CDM and
  geometry-driven Milky Way rotation curve models with Gaia DR2},'' {\em Mon.
  Not. Roy. Astron. Soc.}, vol.~496, no.~2, pp.~2107--2122, 2020.

\bibitem{1972ApJ}
J.~M. {Bardeen}, W.~H. {Press}, and S.~A. {Teukolsky}, ``{Rotating Black Holes:
  Locally Nonrotating Frames, Energy Extraction, and Scalar Synchrotron
  Radiation},'' {\em apj}, vol.~178, pp.~347--370, Dec. 1972.

\bibitem{Ludwig:2021kea}
G.~O. Ludwig, ``{Galactic rotation curve and dark matter according to
  gravitomagnetism},'' {\em Eur. Phys. J. C}, vol.~81, no.~2, p.~186, 2021.

\bibitem{Ruggiero:2021lpf}
M.~L. Ruggiero, A.~Ortolan, and C.~C. Speake, ``{Galactic Dynamics in General
  Relativity: the Role of Gravitomagnetism},'' 12 2021.
\newblock arXiv:2112.08290.

\bibitem{Bini:2008uyx}
D.~Bini, C.~Cherubini, A.~Geralico, and R.~T. Jantzen, ``{Physical frames along
  circular orbits in stationary axisymmetric spacetimes},'' {\em Gen. Rel.
  Grav.}, vol.~40, p.~985, 2008.

\bibitem{Winicour}
J.~Winicour, ``All stationary axisymmetric rotating dust metrics,'' {\em
  Journal of Mathematical Physics}, vol.~16, no.~9, pp.~1806--1808, 1975.

\bibitem{Synge}
J.~L. SYNGE, ``Group motions in space-time and doppler effects,'' {\em Nature},
  vol.~198, no.~4881, pp.~679--679, 1963.

\bibitem{Ruggiero:2002hz}
M.~L. Ruggiero and A.~Tartaglia, ``{Gravitomagnetic effects},'' {\em Nuovo
  Cim.}, vol.~B117, pp.~743--768, 2002.

\bibitem{Mashhoon:2003ax}
B.~Mashhoon, ``{Gravitoelectromagnetism: A Brief review},'' 11 2003.

\bibitem{Misner:1974qy}
C.~W. Misner, K.~S. Thorne, and J.~A. Wheeler, {\em {Gravitation}}.
\newblock San Francisco: W. H. Freeman, 1973.

\bibitem{1965ApJ...142.1488C}
S.~{Chandrasekhar}, ``{The Post-Newtonian Equations of Hydrodynamics in General
  Relativity.},'' {\em apj}, vol.~142, p.~1488, Nov. 1965.

\bibitem{Bonazzola1973TheVT}
S.~Bonazzola, ``The virial theorem in general relativity,'' {\em The
  Astrophysical Journal}, vol.~182, p.~335, 1973.

\bibitem{Astesiano:2021ren}
D.~Astesiano, S.~L. Cacciatori, and F.~Re, ``{Towards a full general
  relativistic approach to galaxies},'' 6 2021.
\newblock arXiv:2106.12818.

\bibitem{Mashhoon:1999nr}
B.~Mashhoon, F.~Gronwald, and H.~I.~M. Lichtenegger, ``{Gravitomagnetism and
  the clock effect},'' {\em Lect. Notes Phys.}, vol.~562, pp.~83--108, 2001.

\end{thebibliography}

\end{document}